\documentclass[prb,reprint, superscriptaddress]{revtex4-1}

\usepackage{graphicx}
\usepackage{amsmath,amsfonts,amssymb}
\usepackage[colorlinks=true,linkcolor=blue,citecolor=blue]{hyperref}

\usepackage{siunitx}
\usepackage{layouts}
\usepackage{todonotes}

\begin{document}


\title{Wideband and on-chip excitation for dynamical spin injection into graphene} 



\author{David I. Indolese}
\thanks{These authors contributed equally.}
\affiliation{Department of Physics, University of Basel, Klingelbergstrasse 82, CH-4056 Basel, Switzerland}

\author{Simon Zihlmann}
\thanks{These authors contributed equally.}
\affiliation{Department of Physics, University of Basel, Klingelbergstrasse 82, CH-4056 Basel, Switzerland}

\author{P\'eter Makk}
\email{peter.makk@unibas.ch}
\affiliation{Department of Physics, University of Basel, Klingelbergstrasse 82, CH-4056 Basel, Switzerland}
\affiliation{Department of Physics, Budapest University of Technology and Economics and Nanoelectronics 'Momentum'
Research Group of the Hungarian Academy of Sciences, Budafoki ut 8, 1111 Budapest, Hungary}

\author{Christian J\"unger}
\affiliation{Department of Physics, University of Basel, Klingelbergstrasse 82, CH-4056 Basel, Switzerland}

\author{Kishan Thodkar}
\affiliation{Department of Physics, University of Basel, Klingelbergstrasse 82, CH-4056 Basel, Switzerland}

\author{Christian Sch\"onenberger}
\affiliation{Department of Physics, University of Basel, Klingelbergstrasse 82, CH-4056 Basel, Switzerland}



\date{\today}

\begin{abstract}
Graphene is an ideal material for spin transport as very long spin relaxation times and lengths can be achieved even at room temperature. However, electrical spin injection is challenging due to the conductivity mismatch problem. Spin pumping driven by ferromagnetic resonance is a neat way to circumvent this problem as it produces a pure spin current in the absence of a charge current. Here, we show spin pumping into single layer graphene in micron scale devices. A broadband on-chip RF current line is used to bring micron scale permalloy (Ni$_{80}$Fe$_{20}$) pads to ferromagnetic resonance with a magnetic field tunable resonance condition. At resonance, a spin current is emitted into graphene, which is detected by the inverse spin hall voltage in a close-by platinum electrode. Clear spin current signals are detected down to a power of a few milliwatts over a frequency range of \SIrange{2}{8}{\giga\hertz}. This compact device scheme paves the way for more complex device structures and allows the investigation of novel materials.
\end{abstract}


\maketitle 

\section{Introduction}
\label{sec:intro}
Graphene has proven to be an excellent material for spintronic applications \cite{2014_Han} with spin relaxation times on the order of \SI{10}{\nano\second} \cite{2016_Droegeler} and spin relaxation lengths of \SI{24}{\micro\metre} \cite{2015_Ingla-Aynes} at room temperature. However, the conductivity mismatch problem \cite{2000_Schmidt} poses severe challenges for efficient electrical spin injection into graphene as it requires a tunnel barrier between the ferromagnetic contact and graphene. Oxide tunnel barriers are the key ingredient for achieving large spin signals in magnetic tunnel junctions \cite{1995_Miyazaki, 1995_Moodera}. However, high quality oxide tunnel barriers are hard to grow on 2D materials (e.g. graphene) \cite{2008_Wang, 2008_Wang_a}. Insulating or semiconducting 2D materials have been investigated as possible tunnel barriers \cite{2014_Friedman, 2017_Omar} and hexagonal boron nitride has proven to be particularly useful \cite{2013_Yamaguchi, 2014_Fu, 2014_Kamalakar, 2016_Singh, 2017_Gurram, 2018_Gurram}.

Spin pumping driven by ferromagnetic resonance (FMR) is a another way to circumvent the conductivity mismatch as it produces a pure spin current in the absence of a charge current \cite{2002_Tserkovnyak, 2002_Brataas, 2011_Ando_a}. The emission of a pure spin current goes along with an enhanced damping of the FMR \cite{2002_Tserkovnyak}.

An enhanced damping of the FMR has been observed in metallic structures \cite{1979_Silsbee, 2002_Mizukami} as well as in graphene based devices \cite{2012_Patra}. Compared to metallic structures, graphene has the advantage that its properties are gate tunable. It has been shown theoretically that this is also the case for the spin mixing conductance \cite{2016_Inoue, 2015_Rahimi}, which describes the spin pumping efficiency.

Even though first hints on spin pumping into graphene have been observed, the detection of a spin current in graphene was still missing until recently. Tang et al. \cite{2013_Tang} showed spin pumping and the detection of a spin current by the inverse spin Hall effect in palladium \cite{2013_Tang}. However, these experiments with macroscopic samples were performed in a RF cavity and therefore at a fixed frequency and at high power levels (on the order of \SI{100}{\milli\watt}).

In this letter, we show that the implementation of an on-chip RF current line to locally excite micron scale permalloy (Py, Ni$_{80}$Fe$_{20}$) pads comes with the advantage of a compact sample design, which in addition allows for: 1) broad frequency range of operation, 2) low RF power levels. Additionally, we detect spin currents using the inverse spin Hall effect in a platinum electrode.

\section{Working principle}
\label{sec:working_principle}
\begin{figure}[htbp]
	\centering
	\includegraphics[width=8.5cm]{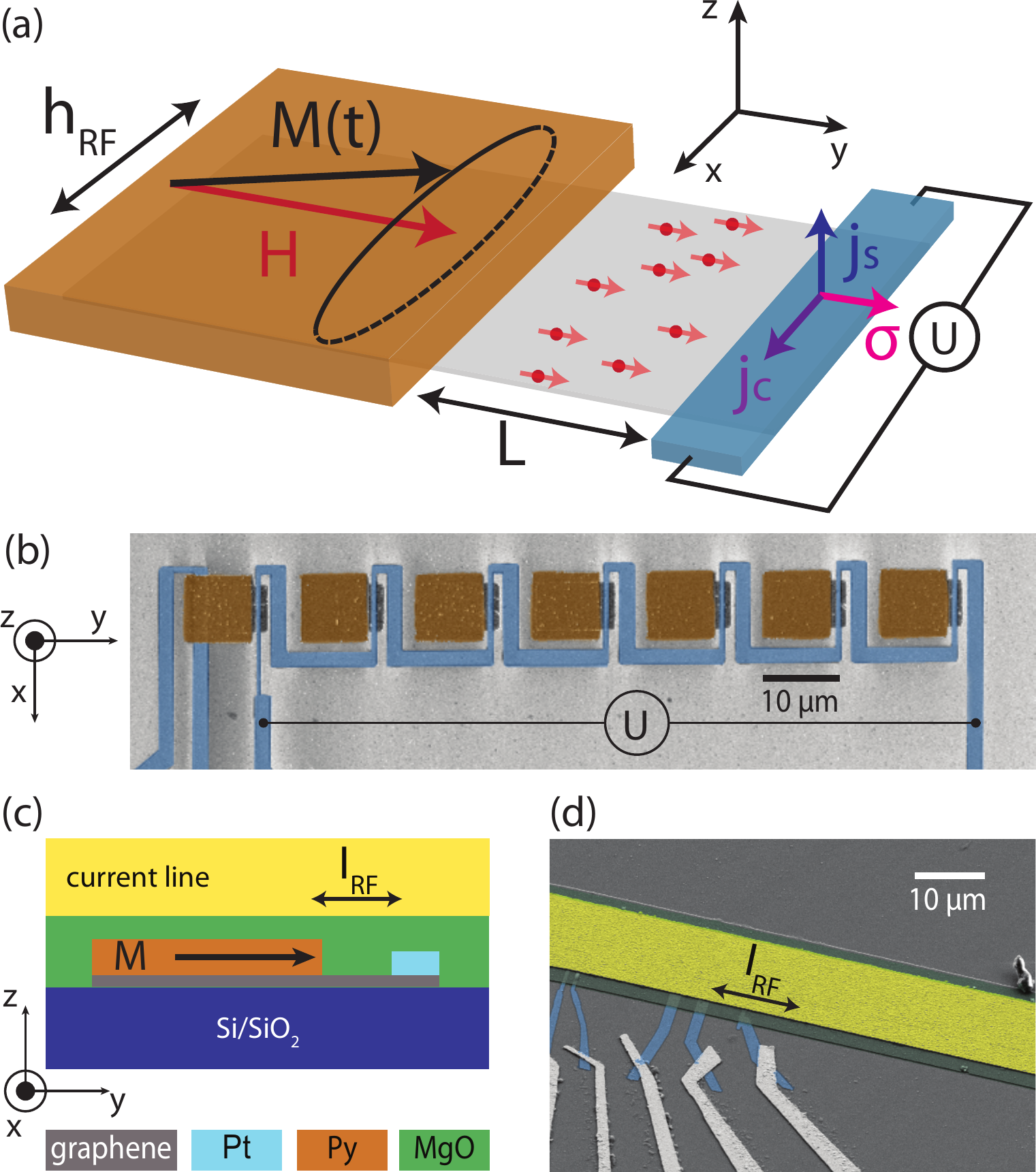}
	\caption{\label{fig:device}\textbf{Device schematics:} \textbf{(a)} shows a schematic drawing of the device with the ferromagnetic Py pad on the left (brown), the graphene spin transport channel in the middle (grey) and the Pt strip on the right (blue) that is used to convert the spin current into a voltage using the inverse spin-Hall effect. \textbf{(b)} shows a false-colour scanning electron micrograph of seven devices. The respective Pt detection strips are connected in series. On top of the devices a RF current line is fabricated to drive the FMR of the Py pads. This is seen in a schematic cross cross section in \textbf{(c)} and as a top view in \textbf{(d)}.}
\end{figure}

A schematic drawing of the investigated samples is shown in Fig.~\ref{fig:device}~(a). On the left a spin current is dynamically injected into the graphene spin transport channel, whereas on the right (at a distance $L$) the spin current is converted into a charge current using a platinum electrode employing the inverse spin-Hall effect (ISHE) \cite{2015_Sinova, 2017_Yan, 2017_Torres}. An external magnetic field H (red arrow along the y-direction) defines the equilibrium magnetization and a RF magnetic field h$_\mathrm{rf}$ (black double arrow along the x-direction) is used to resonantly drive the magnetization M(t) (black arrow) leading to spin pumping with the injection of a spin current into graphene. The DC component of the spin current $j_s$, with a spin orientation $\sigma$ parallel to the external magnetic field H, is converted into a charge current $j_c$ in a platinum electrode as a result of the inverse spin Hall effect \cite{2015_Sinova, 2017_Yan, 2017_Torres}.

All samples presented here were fabricated on intrinsic, high resistive silicon wafers (with \SI{170}{\nano\metre} of SiO$_2$ on top) to reduce RF losses. In a first step CVD graphene \cite{2017_Thodkar} was transferred by a conventional wet transfer method using PMMA as a supporting layer and ammonium persulfate as the copper etchant. The graphene sheet was then patterned into an array of rectangles with width of \SI{8}{\micro\metre} and length of \SI{12}{\micro\meter} by e-beam lithography and reactive ion etching. Next, the platinum electrodes were deposited either by sputter deposition or by thermal evaporation. The thickness of the Pt electrodes was kept at a maximum of $d_{Pt}$~\SI{\approx10}{\nano\metre} to avoid shunting effects that occur for Pt electrodes much thicker than the spin relaxation length within Pt ($\lambda_{Pt}$). Seven devices were connected in series employing a meander structure of the Pt electrode to increase the total signal, see Fig.\ref{fig:device}~(b). For a clean fabrication of the ferromagnetic permalloy structures, a fabrication procedure based on ZEP resist was used \cite{2014_Samm}. Py pads of \SI{8x8}{\micro\metre} were patterned on top of the graphene, see also Fig.~\ref{fig:device}~(b) for a false-colour electron micrograph of a sample at this fabrication stage. A thin Py layer together with the negligible crystal anisotropy of Py favours a homogeneous in-plane magnetization aligned with the external magnetic field H \cite{2004_Hacia, 2010_Aurich}. The length of the spin transport channel $L$ is defined by the Py and the Pt electrode separation.

A layer of MgO (\SI{75}{\nano\metre} thick) was used to insulate the device from the RF current line, which was fabricated directly on top of the sample to maximize the amplitude of h$_{RF}$ driving magnetic field, see Fig.~\ref{fig:device}~(c) for a cross section and Fig.~\ref{fig:device}~(d) for a false-colour electron micrograph. The RF current line consists of \SI{5}{\nano\metre} Ti and \SI{100}{\nano\metre} of Cu with \SI{45}{\nano\metre} of Au on top that prevents the copper from oxidation.

Here, a total of three devices are discussed. Two of which contain a graphene spin transport channel (sample A and B) whereas device C serves as a reference device without graphene.

\section{Results}
\label{sec:results}
A vector network analyser was used to detect the FMR of the Py pads. This is achieved by sourcing a RF signal that is sent into a \SI{50}{\ohm} transmission line on a circuit board that is terminated by the RF current line on the sample chip, which acts as a short. At the same time, the reflection coefficient S$_{11}$ is used to measure the resonance condition. The voltage $U$ at the Pt electrode was measured with a lock-in technique employing magnetic field modulation ($\mu_0dH \sim$\SI{2.75}{\milli\tesla}) at a frequency of \SI{377}{\hertz}. This technique has the advantage that it is more sensitive compared to DC measurements and it is not affected by thermal voltages that can drift over the long time scales of the measurements.
	
\subsection{Ferromagnetic resonance}
\label{subsec:FMR}
The magnetization dynamics of a classical macro spin $\vec{M}$ in an effective magnetic field $\vec{H}_{eff}$ is described by the Landau-Lifshitz-Gilbert equation \cite{1980_Landau, 1955_Gilbert}
\begin{equation}
	\label{eq:LLG}
	\frac{d\vec{M}}{dt} = \gamma\vec{H}_{eff}\times\vec{M} + \alpha\vec{M}\times\frac{d\vec{M}}{dt},
\end{equation}
where $\gamma = g\mu_B/\hbar$ is the gyromagnetic ratio with $g$ the Land\'e g-factor and $\mu_B = e\hbar/(2m_e)$ the Bohr magneton, where $\hbar$ is the reduced Planck constant, and $\alpha$ the Gilbert damping constant. In the case of a thin film with the external magnetic field applied in-plane ($H_{eff}\sim H_{ext}$), the resonance condition is given by the Kittel formula \cite{1996_Kittel}:
\begin{equation}
	\label{eq:spin_pumping_Kittel}
	f_{res} = \frac{\gamma\mu_0}{2\pi}\sqrt{H_{ext}\left(H_{ext} + M_s\right)},
\end{equation}
where $\mu_0$ is the vacuum permeability and $M_s$ is the saturation magnetization. 

\begin{figure}[htbp]
	\includegraphics[scale=1]{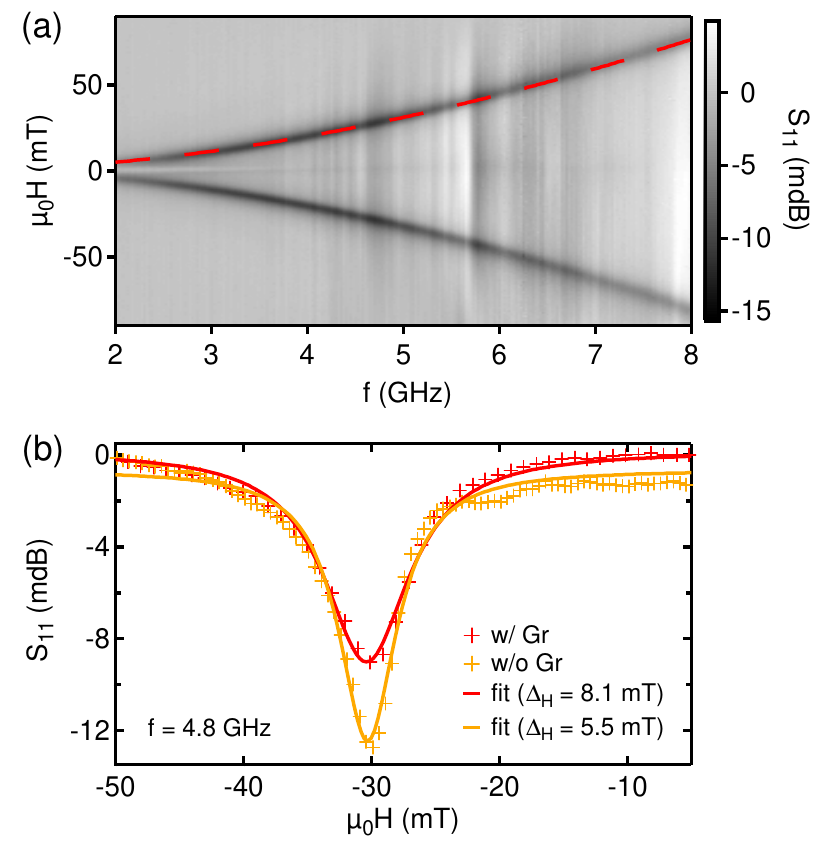}
	\caption{\label{fig:FMR}\textbf{Ferromagnetic resonance condition:} (a) shows S$_{11}$ as a function of magnetic field and frequency with a frequency dependent background subtracted. (b) shows $S_{11}$ as a function of magnetic field at f~=~$\SI{4.8}{\giga\hertz}$ for a device without graphene (orange) and for a device with graphene (red). Lorentzian fits are used to deduce $\Delta_H$ as indicated in the legend.}
\end{figure}

Fig.~\ref{fig:FMR}~(a) shows S$_{11}$ as a function of external magnetic field and frequency measured on sample A. In order to eliminate the standing wave background due to reflections at each RF connector, a frequency dependent background was subtracted (\SI{100}{\milli\tesla} trace). The remaining vertical lines originate from a weak magnetic dependence of the standing wave background. At the FMR condition RF power is absorbed by the precessing magnetization. This is seen in S$_{11}$, which is reduced at the resonance since not all of the RF signal is reflected at the terminating short of the RF current line. The saturation magnetization $\mu_0M_s=$~\SI{0.96}{\tesla} was extracted by fitting the resonance position with Eq.~\ref{eq:spin_pumping_Kittel}, while fixing $g=2$ to literature values \cite{2006_Costache}. The extracted $M_s$ agrees well with literature values \cite{2006_Costache, 2007_Chen}.

The line width of the FMR is given by the damping term $\alpha$ in Eq.~\ref{eq:LLG}. Cuts at $f=$~\SI{4.8}{\giga\hertz} are shown in Fig.~\ref{fig:FMR}~(b) for sample A and for sample C (with and without graphene). The full width at half maximum ($\Delta_H$) was extracted by fitting a Lorentzian to the data. A significant larger $\Delta_H$ was observed for the sample with graphene indicating an additional damping term. The linewidth of the FMR, $\Delta_H$, can be related to the Gilbert damping \cite{2010_Mosendz_b}
\begin{equation}
	\label{eq:spin_pumping_linewidth}
	\Delta_H = \frac{4\pi\alpha}{\gamma}f_{res}.
\end{equation}
Here, inhomogeneous sample-dependent broadening of the linewidth was neglected since it was shown \cite{2012_Patra} and also found in our measurements to be negligible. This additional damping term can be interpreted as spin pumping into graphene \cite{2002_Tserkovnyak} and the difference in linewidth of a sample with graphene ($\Delta_{H, Py/Gr}$) and one without graphene ($\Delta_{H, Py}$) can be used to estimate the real part of the effective spin-mixing conductance \cite{2002_Tserkovnyak_a, 2010_Mosendz_b}
\begin{equation}
	\label{eq:spin_pumping_mixing_conductance}
	g_{\uparrow\downarrow} = \frac{M_sd_{Py}}{\hbar f}\left(\Delta_{H, Py/Gr} - \Delta_{H, Py}\right),
\end{equation}
where $d_{FM}$ is the thickness of the ferromagnetic Py layer. The imaginary part of the spin mixing conductance can be neglected since it is much smaller than the real part for metallic ferromagnets \cite{2002_Tserkovnyak_a}. The spin-mixing conductance is a measure of the efficiency of the spin injection and was here estimated to be \SI{\sim 1e20}{\per\square\metre} using $\mu_0M_s=$~\SI{0.96}{\tesla} as extracted above, $d_{FM} = $~\SI{30}{\nano\metre} and a literature value of $g=2$ \cite{2006_Costache}. This values is roughly a factor of two larger than previously reported in similar Py/graphene systems \cite{2012_Patra}.

\subsection{Spin current and inverse spin Hall voltage}
\label{subsec:ISHE_voltage}
The DC component of the spin current density flowing across the Py/graphene interface in z-direction due to spin pumping is given by 
\begin{equation}
	\label{eq:spin_current}
	j_s = P\cdot\frac{hf}{4\pi}g_{\uparrow\downarrow}\cdot\sin^2(\theta),
\end{equation}
where $P$ is a correction factor that accounts for a non-circular precession mode and $\theta$ is the precession angle of the magnetization around the effective field $H_{eff}$ \cite{2010_Mosendz}. We estimate the correction factor for the non-circular precession at \SI{5}{\giga\hertz} to be around 0.6 based on Ref.~\onlinecite{2010_Mosendz_b}. Eq.~\ref{eq:spin_current} describes the spin current density across the Py/Gr interface. We would like to note that our device geometry differs from the conventional metallic bilayer structures since a spin current is detected laterally. Therefore, only a region of length $\lambda_{Py}$ at the Py interface to the lateral graphene spin transport channel can contribute to a spin current that is detected at a distance L.

The spin current due to spin pumping is detected with a Pt electrode placed at a distance $L$ (\SI{600}{\nano\metre} for device A and \SI{700}{\nano\metre} for device B) from the Py pad, see also Fig.~\ref{fig:device}. The charge current $j_c$ due to the inverse spin-Hall effect can be detected as a voltage in an open-circuit configuration. This voltage changes sign if the direction of the external magnetic field is reversed as the spin polarization $\sigma$ is parallel to the external magnetic field ($j_c\sim \sigma\times j_s$ \cite{2015_Sinova}). The voltage due to the inverse spin-Hall effect follows the line shape of the FMR and is therefore described by a Lorentzian.

\begin{figure}[htbp]
	\centering
	\includegraphics[scale=1]{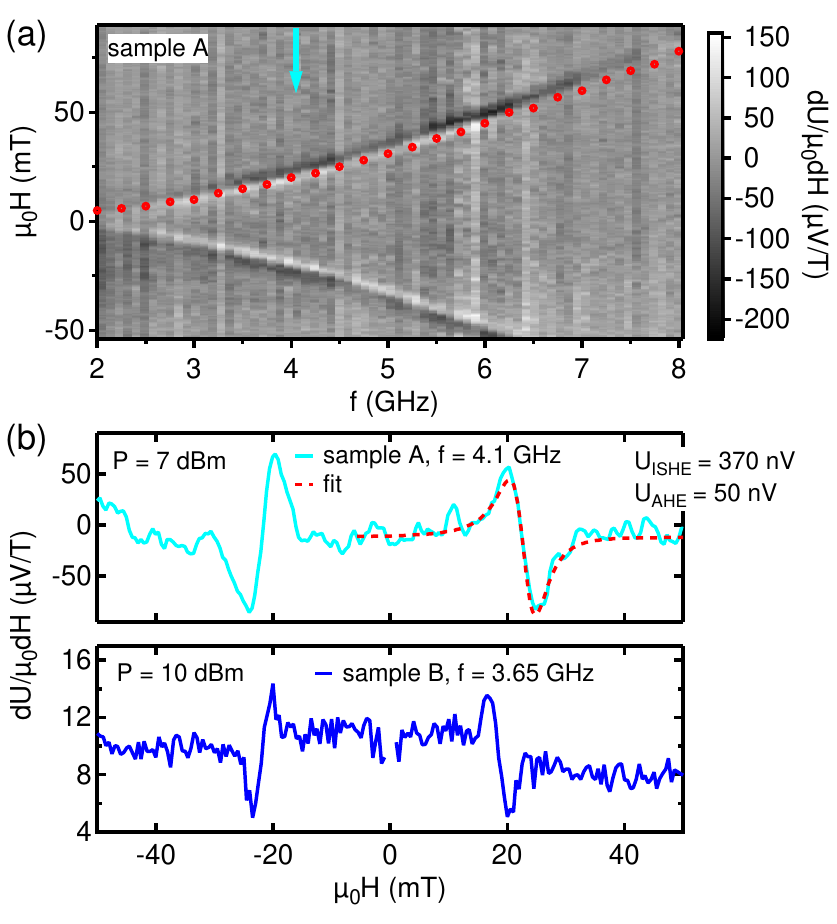}
	\caption{\label{fig:ISH_f}\textbf{Inverse spin Hall voltage at Pt electrode:} (a) shows the derivative of the voltage measured at the Pt electrode with respect to the magnetic field as a function of magnetic field and frequency for sample A. The superimposed red dots mark the position of the FMR condition extracted from a measurement of S$_{11}$ of a different sample. (b) shows the cut indicated in (a) and a cut from sample B. The signal clearly shows the mirror symmetry with respect to zero magnetic field. The red dashed line is a fit with Eq.~\ref{eq:spin_pumping_ISHE_fitting}. In the case of sample B, the data points around zero magnetic field were removed due to technical limitations.}
\end{figure}

Here, the voltage $U$ at the Pt electrode was measured with a lock-in technique employing magnetic field modulation. Therefore, we recorded $dU/\mu_0dH$ as a function of RF frequency and magnetic field as shown in Fig.~\ref{fig:ISH_f}~(a). The signal follows the FMR condition, which is indicated by red dots. The slight discrepancy at larger frequencies can be explained by sample to sample variation as the FMR condition was extracted from a different sample but with nominally equal design. Fig.~\ref{fig:ISH_f}~(b) shows $dU/\mu_0dH$ as a function of magnetic field and reveals the expected lineshape of a derivative Lorentzian. Since $U$ depends on the spin orientation $\sigma$, which itself depends on the magnetic field $H$, it has opposite sign for $H_{FMR}$ and $-H_{FMR}$, where $H_{FMR}$ is the magnetic field at which the FMR occurs. As a consequence, $dU/\mu_0dH$ shows a dip-peak structure for negative values of $H$ and a peak-dip structure for positive values of $H$. This behaviour is seen for all frequencies investigated here, see Fig.~\ref{fig:ISH_f}~(a). Similar results were obtained for sample B, shown in Fig.~\ref{fig:ISH_f}~(c).

As motivated above, a magnetic field modulation based measurement technique has its advantages when it comes to sensitivity and influences by spurious effects. However, it can itself lead to a background signal. The small modulation of the magnetic field induces a voltage in the wires connecting the sample to the voltage amplifiers due to simple inductive pick-up. This voltage is magnetic field dependent as it scales with the modulation amplitude that itself depends on the magnetic field due to a non-linear current to field conversion of the magnet set-up. In order to remove this background, the voltage at the Pt electrode was once measured with the microwave source turned on and once with the microwave source turned off. The difference of these two measurements is shown in Fig.~\ref{fig:ISH_f} and is used in the following analysis.

The measurement set-up presented above is only sensitive to voltages that develop in x-direction. Contributions due to the anomalous Hall effect (AHE) can therefore be expected since RF eddy currents are induced by the RF magnetic field in the Py pads. These currents flow in the yz-plane in the permalloy and in combination with a varying magnetization in the xz-plane an anomalous Hall voltage can be expected to appear. This voltage will consist of a component at twice the frequency and of a down mixed DC component along the x direction and therefore, we include the AHE into the analysis. In order to separate the ISHE from the AHE, the measured $dU/\mu_0dH$ was fitted with:

\begin{equation}
	\label{eq:spin_pumping_ISHE_fitting}
	\begin{split}
	U(H) = U_{ISHE}\frac{\Delta_H^2}{\left(H-H_{FMR}\right)^2 +	\Delta_H^2} +\\ U_{AHE}\frac{-2\Delta_H \left( H- H_{FMR}\right)}{\left( H-H_{FMR}\right)^2 + \Delta_H^2},
	\end{split}
\end{equation}
that captures both contributions \cite{2006_Saitoh}. Here, $U_{ISHE}$ and $U_{AHE}$ represent the amplitudes of the contribution of the ISHE and the AHE. The ISHE contribution follows the Lorentzian shape of the FMR condition and has its maximum at the resonance frequency. On the other hand, the AHE contribution displays a different line shape with a contribution that changes sign across the resonance conditions since $M(t)$ phase shifts by $\pi$ at resonance. Therefore, the ISHE and the AHE contribution can be disentangled by their spectral shape \cite{2007_Inoue, 2006_Saitoh}.

\subsection{Power dependence}
\label{subsec:ISHE_power_dep}

\begin{figure}[htbp]
	\centering
	\includegraphics[scale=1]{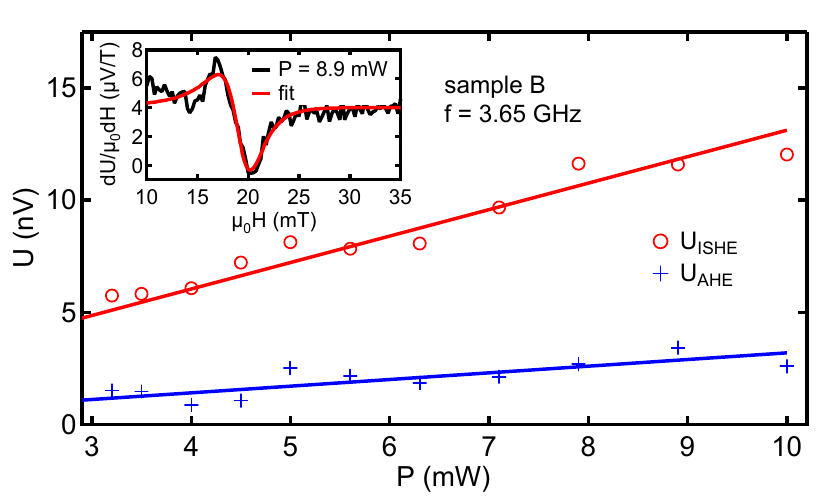}
	\caption{\label{fig:ISH_power}\textbf{Power dependence of the voltage at the Pt electrode:} The contribution of the ISHE and the AHE to the voltage at the Pt electrode are shown as a function of microwave power. The inset shows an actual measurement with a fit to Eq.~\ref{eq:spin_pumping_ISHE_fitting}. Solid lines are linear guides to the eye.}
\end{figure}

The data and the fit with the derivative of Eq.~\ref{eq:spin_pumping_ISHE_fitting} is shown in Fig.~\ref{fig:ISH_f}~(b) for sample A and in the inset of Fig.~\ref{fig:ISH_power} for sample B. Power dependence was investigated on sample B as shown in Fig.~\ref{fig:ISH_power}, where the ISHE and the AHE contribution are shown separately. The contribution due to the ISHE is much larger than the contribution due to the AHE for any microwave power investigated.

$U_{SHE}$ linearly depends on the spin current density $j_s \sim \sin^2\left(\theta\right)$. The cone angle $\theta \sim 2h_{RF}/\Delta_H$ itself is linearly depending on the driving RF field $h_{RF}$ \cite{2007_Guan}. For small angles this leads to a linear power dependence of $U_{SHE} \sim h_{RF}^2 \sim \sqrt{P}^2 \sim P$ since $h_{RF}$ scales with the square root of the applied power. The power dependence of $U_{SHE}$ shown in Fig.~\ref{fig:ISH_power} is consistent with the linear dependence as indicated by the red solid line.

$U_{AHE}$ depends linearly on the eddy currents that scale linearly with $h_{RF}$ and it also depends linearly on the precessing magnetization $M(t)$, which also scales linearly with $h_{RF}$. Therefore, $U_{AHE}\sim h_{RF}^2 \sim \sqrt{P}^2 \sim P$ and a linear power dependence results. The AHE contribution indeed scales linearly with RF power as one can see in Fig.~\ref{fig:ISH_power}.

\section{Discussion and interpretation}
\label{sec:discussion}
The observed broadening of the FMR linewidth when the FM is in contact with the graphene sheet shows the presence of an additional damping channel that can be explained by spin pumping into graphene. The two times larger spin mixing conductance extracted here compared to the literature values \cite{2012_Patra, 2013_Tang} might be explained by the cleaner ZEP based fabrication protocol. Successful dynamical spin injection into graphene is further supported by the observation of a voltage at the Pt electrode that follows a derivative Lorentzian lineshape that is expected for the ISHE. This voltage follows the FMR condition over a broad frequency range and shows the sign change for negative magnetic field. Power dependence of this voltage reveals a linear scaling of the ISHE contribution as expected, with a minor contribution from the AHE.

\subsection{Quantitative analysis}
A quantitative analysis of the injected spin current compared with the inverse spin-Hall voltage is given in the following.

The injected spin current density at the Py/graphene interface can be estimated from eq.~\ref{eq:spin_current}. According to Guan et al.~\cite{2007_Guan}, the precession angle at resonance is given by $\theta \sim 2h_{RF}/\Delta_H\sim$~\SI{8.4}{\degree}. The driving RF magnetic field $h_{RF}$ was estimated to be \SI{0.6}{\milli\tesla} using the Biot-Savart law where the RF current is assumed to flow homogeneously within the RF current line and where the RF current is given by the applied power of \SI{7}{dBm} over a \SI{50}{\ohm} impedance. This is the maximum h$_{RF}$ that can be expected since neither RF losses in the cables nor reflections at the connectors are included.

We then estimate $j_s\sim$~\SI{2.8e-7}{\joule\per\square\metre} at \SI{4.1}{\giga\hertz} and \SI{7}{dBm}. In order to get the lateral spin current (in y-direction), one has to determine the area that contributes to the injected current. Since the Py is very well coupled to the graphene below, only a narrow strip of the size $w_{Gr/Py}\times \lambda_{Py}$ will contribute as in the other parts the spins will have relaxed before reaching the spin transport channel. This then results in a lateral spin current density of $j_s^y = j_s \cdot \lambda_{Py}\sim$~\SI{1.2e-15}{\joule\per\metre}, using $\lambda_{Py}=$~\SI{4.3}{\nano\metre} \cite{1999_Dubois}. Similar spin current densities are commonly realized with electrical spin injection with tunnel coupled ferromagnetic contacts.

Using this current density now we can calculate the inverse spin Hall voltage appearing on the Pt electrode and we can compare it with the measured values. The measured voltage U at the Pt electrode is given by:
\begin{equation}
	\label{eq:U_ISHE}
	\begin{split}
	U_{ISHE} = \frac{2e}{\hbar}\alpha_{Pt}\rho_{Pt} w_{gr/Pt}\frac{j_s^y}{l_{Pt}}\\ \cdot \frac{\lambda_{Pt}}{d_{Pt}}\frac{1-\exp\left(-d_{Pt}/\lambda_{Pt}\right)}{1+\exp\left(-d_{Pt}/\lambda_{Pt}\right)}\cdot \frac{\lambda_Pt}{l_{Pt}},
	\end{split}
\end{equation}
where the first term describes the spin to charge conversion via the inverse spin Hall effect, the second term incorporates spin relaxation in z-direction within the Pt electrode following Ref.~\onlinecite{2017_Torres} and the last term is the extension of that model considering also spin relaxation in y-direction and shunting due to the metallic electrode. Note that $l_{Pt}\gg\lambda_{Pt}$ and therefore the exponential corrections can be neglected in y-direction. We used $\rho_{Pt}\sim$~\SI{46}{\micro\ohm\centi\metre}, $w_{gr/Pt} = 7\times$~\SI{8}{\micro\metre}~=~\SI{56}{\micro\metre} (seven devices, each \SI{8}{\micro\metre}), $l_{Pt} = $\SI{400}{\nano\metre} and $d_{Pt} = $\SI{10}{\nano\metre} that were determined experimentally, whereas the spin Hall angle $\alpha_{Pt}\sim$~0.15 \cite{2017_Torres} and $\lambda_{Pt}\sim$~\SI{5}{\nano\metre} \cite{2015_Sinova} were taken from literature. A lateral spin current density $j_s^y\sim$~\SI{1.2e-15}{\joule\per\metre} leads to an expected voltage $U_{ISHE}\sim $~\SI{170}{\nano\volt}, which is within a factor of two from the experimentally determined value.

The difference between the measured voltage and expected voltage at the platinum electrode can have serveral origins. First, the spin relaxation within the graphene spin transport channel is neglected since spin relaxations lengths of the order of \SI{1}{\micro\metre} are commonly obtained also for low quality graphene devices. Next, we would like to note that the cone angle is only a rough and upper estimate and an experimental determination would reduce the uncertainty of that value. Moreover, several parameters used for the estimation of the voltage at the platinum electrode are not well known and especially a large spread of the values for the spin Hall angle of Pt is found in literature \cite{2015_Sinova}.

\section{Conclusion}
\label{sec:conclusion}
The development of the compact, on-chip and broadband excitation scheme for spin pumping into graphene and the detection of a spin current with a Pt electrode paves the way for future studies focussing on the extraction of spin transport parameters. Hanle measurements, as shown in spin pumping experiments in silicon \cite{2015_Pu}, could be performed in vector magnet set-ups, where an additional magnetic field in the z-direction is available. Another step forward would be the implementation of ferromagnetic strips instead of squares, which would allow spin pumping experiments at zero external in-plane magnetic field due to the non-zero remanent magnetization of nanomagnets \cite{2006_Costache, 2006_Costache_a}.

In conclusion, we demonstrated that FMR can be observed in micronscale Py/graphene heterostructures with on-chip and wideband microwave excitation in a simple reflection measurement. The increased damping of the FMR in Py pads connected to graphene suggested the presence of spin-pumping, which is further supported by the detection of a spin current at a Pt electrode employing the inverse spin Hall effect. This direct and compact way of spin pumping into graphene to power levels as low as \SI{3}{\milli\watt} paves the way for further studies on the spin dynamics in graphene and related heterostructures.

Future studies could investigate the spin-to-charge conversion in graphene itself as recently reported by Mendes et al. \cite{2015_Mendes}. In addition, heterostrucutres of graphene and transition metal dichalcogenides have shown a greatly enhanced spin-orbit coupling \cite{2016_Wang} with a dominating valley-Zeeman term \cite{2018_Zihlmann}. These systems are expected to show large spin-Hall angles \cite{2017_Garcia} that would allow for an even more efficient spin-to-charge conversion. This is especially interesting and important since graphene is a promising candidate for future building blocks in spintronic applications (e.g. spin torque nano oscillators) considering that it can withstand large current densities \cite{2007_Moser} and large spin accumulations \cite{2017_Gurram} can be achieved.


%
%

%

\begin{acknowledgments}
This work has received funding from the European Union’s Horizon 2020 research and innovation programme under grant agreement No 696656 (Graphene Flagship), the Swiss National Science Foundation, the Swiss Nanoscience Institute, the Swiss NCCR QSIT and ISpinText FlagERA network OTKA PD-121052, OTKA FK-123894 and OTKA  K112918. This research was supported by the National Research, Development and Innovation Fund of Hungary within the Quantum Technology National Excellence Program (Project Nr.  2017-1.2.1-NKP-2017-00001). P.M. acknowledges support as a Bolyai Fellow.
\end{acknowledgments}

%

\end{document}